\documentstyle[12pt,aasms4,psfig]{article}

\newcommand {\ie} {{\it  i.e.}}

\newcommand {\eg} {{\it e.g.}}
\newcommand {\etal} {{\it et~al.}}

\newcommand {\go} {\mathrel{\hbox{\rlap{\lower.55ex \hbox {$\sim$}}
        \kern-.3em \raise.4ex \hbox{$>$}}}}
\newcommand {\lo} {\mathrel{\hbox{\rlap{\lower.55ex \hbox {$\sim$}}
        \kern-.3em \raise.4ex \hbox{$<$}}}}    
\def\Msun{\ifmmode M_\odot \else $M_\odot$\fi} 
\def\Mdot{\ifmmode \dot M \else $\dot M$\fi} 
\def\psnb{\ifmmode n_{PSNB} \else $n_{PSNB}$\fi} 
\def\lmxb{\ifmmode n_{LMXB} \else $n_{LMXB}$\fi} 
\def\mrp{\ifmmode n_{MRP} \else $n_{MRP}$\fi} 
\def\taus{\ifmmode \tau_{SFR} \else $\tau_{SFR}$\fi} 
\def\taup{\ifmmode \tau_{PSNB} \else $\tau_{PSNB}$\fi} 
\def\tauel{\ifmmode \tau_{LMXB} \else $\tau_{LMXB}$\fi} 
\def\taum{\ifmmode \tau_{MRP} \else $\tau_{MRP}$\fi} 
\def\ratem{\ifmmode {n_{MRP}\over\tau_{MRP}} \else 
${n_{MRP}\over\tau_{MRP}}$\fi} 
\def\ratel{\ifmmode {n_{LMXB}\over\tau_{LMXB}} \else 
${n_{LMXB}\over\tau_{LMXB}}$\fi} 

\begin{document}

\title{LOW-MASS X-RAY BINARIES, MILLISECOND RADIO PULSARS,
 AND THE COSMIC STAR FORMATION RATE}
\author{Nicholas E. White \& Pranab Ghosh$^{1,2}$}
\affil{Laboratory for High Energy astrophysics}
\affil{Code 662, NASA Goddard Space Flight Center, Greenbelt, MD 20771}

\received{8th June 1998}
\accepted{2nd July 1998}

\slugcomment{To appear in the Astrophysical Journal Letters}

\lefthead{White \& Ghosh}
\righthead{LMXB, MRP, \& Star Formation Rate}

\vskip 2 truein
$^1$ Senior NAS/NRC Resident Research Associate
 
$^2$ On leave from Tata Institute of Fundamental Research, Bombay 400 005,
India
\vskip 2 truein

\begin{abstract}

We report on the implications of the peak in the cosmic star-formation 
rate (SFR) at redshift $z\approx 1.5$ for the resulting population of 
low-mass X-ray binaries(LMXB) and for 
that of their descendants, the millisecond 
radio pulsars (MRP). Since the evolutionary timescales of LMXBs, their 
progenitors, and their descendants are thought be significant fractions
of the time-interval between the SFR peak and the present epoch, there
is a lag in the turn-on of the LMXB population, with the peak activity
occurring at $z\sim 0.5-1$. The peak in the MRP population is delayed
further, occurring at $z\lo 0.5$. We show that the discrepancy between
the birthrate of LMXBs and MRPs, found under the assumption of a
stead-state SFR, can be resolved for the population as a whole when
the effects of a time-variable SFR are included. A discrepancy may
persist for LMXBs with short orbital periods, although a detailed
population synthesis will be required to confirm this. Further, since
the integrated X-ray luminosity distribution of normal galaxies is
dominated by X-ray binaries, it should show strong luminosity evolution
with redshift. In addition to an enhancement near the peak ($z\approx 
1.5$) of the SFR due to the prompt turn-on of the relatively 
short-lived massive X-ray binaries and young supernova remnants, 
we predict 
a second enhancement by a factor $\sim 10$ at a redshift between 
$\sim 0.5$ and $\sim 1$ due to the delayed turn-on of the LMXB 
population. Deep X-ray observations of galaxies out to $z\approx 1$ 
by AXAF will be able to observe this enhancement, and, by determining 
its shape as a function of redshift, will provide an important new 
method for constraining evolutionary models of X-ray binaries.
     
\end{abstract}


\section{Introduction}

It has recently been shown that the cosmic star formation rate (SFR)
increases with redshift, reaching a peak $\sim 10$ times higher than the 
current rate in the redshift interval $\sim 1-2$ (Lilly \etal\ 1996; 
Madau \etal\ 1997; Madau, Pozzetti \& Dickinson 1998, henceforth MPD;
Madau, della Valle \& Panagia 1998, henceforth MVP). In 
this {\it Letter}, we report interesting implications of this for the 
evolution of LMXBs, for that of their descendants, the millisecond 
radio pulsars (MRP), and also for the well-known LMXB-MRP ``birthrate 
problem'' (see Bhattacharya 1995 and references therein). Similar
considerations of the implications of the cosmic SFR for the evolution
of the cosmic supernova rates have recently been undertaken (MVP). 
In essence, the LMXB-MRP birthrate problem 
stems from the observation that, for estimated numbers $N_{LMXB}
\sim 100$ and $N_{MRP}\sim 10^4$ in our galaxy, and for expected 
lifetimes $\tauel\sim 10^8-10^9$ yr and $\taum\sim 10^9-10^{10}$ yr,
the steady-state birthrate of MRPs, $N_{MRP}\over\taum$, exceeds that
of LMXBs, $N_{LMXB}\over\tauel$, by $\go 10$ (Kulkarni and Narayan 1988,
henceforth KN; Cot\'e \& Pylyser 1991, henceforth CP; Lorimer 1995, 
henceforth L95). Many suggested solutions to the problem include (a)
accretion-induced collapse of a white dwarf to a neutron star, (b)
much shorter values of \tauel~for the short-period LMXBs (\ie, those
with orbital periods $\lo 3$ days, say, following the definition of KN),
due, \eg, to X-ray irradiation of the low-mass companion (Tavani 1991, 
henceforth T91), and (c) the possibility that pulsars can be born with 
low magnetic fields and millisecond periods. 
 
We show here that steady-state arguments do not generally apply to 
evolving LMXB and MRP populations with a time-dependent global SFR 
peaking at $z\approx 1.5$, since this peak propagates through the LMXB 
and MRP populations at smaller redshifts, and causes an enhanced MRP 
population at the present epoch. Indeed, except in special circumstances 
outlined in \S 2, there is no basis for expecting an equality between 
the rates $N_{MRP}\over\taum$ and $N_{LMXB}\over\tauel$. We present 
evolutionary calculations which show that the expected number ratio,
$N_r\equiv{\mrp\over\lmxb}$, and rate ratio, $R_r\equiv{N_{MRP}\over\taum}
/{N_{LMXB}\over\tauel}$, at $z=0$ are in agreement with the currently 
observed values for the whole population. However, there may still be a  
discrepancy for the short-period systems. We consider other observational 
tests of our model and indicate how observations of galaxies in the 
redshift range $0.5-1.0$ by AXAF can constrain models for the 
evolution of X-ray binaries.

\section{Evolution of LMXBs \& MRPs With Variable Star Formation Rate}

The standard evolutionary scenario for the 
majority\footnote{We do not consider
the LMXBs and MRPs found in globular clusters, where tidal capture of neutron 
stars in close encounters with stars causes an excess of LMXBs relative to
the overall Galactic population.} of LMXBs and MRPs 
begins with a primordial binary containing a massive OB and a low-mass star (see, 
\eg, Webbink, Rappaport \& Savonije 1983, henceforth WRS; Kalogera 
\& Webbink, 1996, 1998). The massive star rapidly evolves to the 
point of supernova (SN), resulting in the formation of a post-SN 
binary (PSNB) consisting of a neutron star with a low-mass companion, 
which turns into a LMXB when the latter attains Roche lobe contact, 
either due to nuclear evolution or due to orbital decay by 
gravitational radiation and magnetic braking. This, in turn, produces 
a recycled MRP at the end of mass transfer. We demonstrate here the 
basic effects of a time-variable SFR on the above scenario. In this
introductory work, we confine ourselves to a simple description, 
in which the evolution of the number density of each
species (\psnb~for PSNBs, \lmxb~for LMXBs, and \mrp~for MRPs)     
is described by a timescale which is a given number: $\taup$ for the 
evolution of PSNBs into LMXBs, $\tauel$ for that of LMXBs into 
MRPs, and $\taum$ for that of the MRPs.  In reality, $\taup$ and 
$\tauel$ depend on the binary period, $\taup$ on other evolutionary 
parameters (see below) as well, and $\taum$ on evolutionary parameters 
which seem poorly understood at present. To demonstrate the effects of 
these timescales, we run the evolutionary scheme over the ranges of 
their values suggested in the literature: detailed population-synthesis 
studies are deferred to the future.

The evolutions of populations of PSNBs, LMXBs and MRPs in response to a 
time-dependent star-formation rate $SFR(t)$ are given by:
$${\partial\psnb(t)\over\partial t} = \alpha SFR(t)
-{\psnb(t)\over\taup}\,,\eqno(1)$$
$${\partial\lmxb(t)\over\partial t} = {\psnb(t)\over\taup}
-{\lmxb(t)\over\tauel}\,,\eqno(2)$$
$${\partial\mrp(t)\over\partial t} = {\lmxb(t)\over\tauel}
-{\mrp(t)\over\taum}\,.\eqno(3)$$
In equation (1), $SFR(t)$ is that given by MPD and MVP, with the SFR
evolving on a timescale $\taus \approx 6.4\times 10^8$ yr: for all 
calculations reported here, we have used the analytic approximation 
(accurate to within $5\%$) to the SFR given by MVP, which is shown in 
Figure 1. Further, $\alpha$ is a coefficient which determines the rate 
of formation of PSNBs per unit star-formation rate. 
Assuming that the time required by massive newborn 
stars to evolve to the point of supernova is small compared to all other 
evolutionary timescales in the problem, \ie, \taus, \taup, \tauel,
and \taum~(an excellent approximation in view of the value of \taus~ 
given above and the values of \taup, \tauel, and \taum~ 
given below), $\alpha$ is given approximately by $\alpha = {1\over 2}
f_{binary}f_{prim}f_{SN}$. Here, $f_{binary}$ is the fraction of all stars
in binaries, $f_{prim}$ is that fraction of primordial binaries which has
the correct range of stellar masses and orbital periods for evolving into
PSNBs capable of producing LMXBs (KW98), and $f_{SN}$ is that fraction of 
the latter binaries which survives the supernova. The actual value of 
$\alpha$, which sets the overall scale for the sizes of the populations
\psnb, \lmxb~and \mrp~relative to that of the SFR, is irrelevant for 
this study, since we are only interested in the relative sizes of 
\lmxb~and \mrp~here.  

Figure 1 shows the evolution of PSNBs, LMXBs and MRPs described by 
equations (1)-(3): we have displayed our results in terms of the 
redshift $z$, which is related to the cosmic time $t$ by $t_9 = 
13(z+1)^{-3/2}$, where $t_9$ is $t$ in units of $10^9$ yr, and a 
value of $H_0=50$ km s$^{-1}$ Mpc$^{-1}$ has been used (MPD). 
Our choices of representative values of \taup, \tauel, and 
\taum~come from the following considerations. The distribution of 
\taup~with orbital period is rather broad, has a peak in the range 
$1-2\times 10^9$ yr, and is somewhat dependent on supernova 
kick velocity and common-envelope evolution efficiency (Kalogera 1998, 
private communication). The expected range of \tauel~values 
has been discussed extensively in the literature on the birthrate 
problem (KN; CP; T91; L95). The standard WRS mass-transfer time, 
$\tauel\approx 1.1\times 10^9/P_i(d)$ yr has been widely used, where        
$P_i(d)$ is the initial orbital period of the LMXB in days, and, to
ameliorate the problem for short-period LMXBs, it has been suggested that
the effects of X-ray irradiation of the low-mass companion may reduce 
\tauel~to $\sim 10^7$ yr in these systems (T91; L95). For \taum, we used
values based on the compilation by Camilo, Thorsett \& Kulkarni (1994),
which suggests  \taum~in the range $3\times 10^9 - 3\times 10^{10}$ yr.

\section{Discussion}

The results shown in Figures 1(a)-(c)clearly demonstrate that the 
$\sim$ Gyr timescales involved in the evolution of PSNBs, LMXBs and 
MRPs lead to substantial time-lags in the peaks of their populations 
behind the peak in the SFR. The LMXB peak is delayed by several Gyr 
relative to the SFR peak, and appears in the redshift range $z\sim
0.5-1$. The MRP peak is delayed even further, appearing at redshifts
$\lo 0.5$ (including the current epoch). Thus, although the previous 
work of KN and others assumed steady state conditions while comparing
the birthrates of MRP and LMXB, we now see that this assumption is
not correct in general in a universe with time-dependent cosmic SFR.     
There are, however, two limiting cases where this assumption will 
still apply. The first occurs when an asymptotic state is reached for 
all populations, which happens at times much longer than {\it all \/} 
evolutionary timescales in the problem, as illustrated by the case 
shown in Figure 1(d). However, it must be clear that such a situation 
cannot occur within the current age of the universe for any realistic 
choice of \taup, \tauel, and \taum, which is why we had to assume 
unrealistically short timescales for Figure 1(d). This possibility 
is thus of little relevance to the present universe, unless our 
current understanding of LMXB and MRP evolution is completely wrong. 
The second situation obtains when, 
for sufficiently large values of \taum, the 
present epoch (z=0) happens to be at or near the maximum of 
the MRP evolution, where ${\partial\mrp(t)\over\partial t} = 0$ (see 
eq.[3]), so that $\ratem\approx\ratel$ at this epoch. As illustrated
by the case shown in Figure 1(a), such a situation is quite possible 
for realistic values of evolutionary timescales. We have demonstrated 
in Figures 1(b) and (c) that it is also possible have situations in 
which \ratem~is considerably larger than \ratel~in the present epoch 
for plausible values of evolutionary timescales.    

In relating the observational situation to the basic theoretical 
expectations for the number ratio, $N_r$, and rate ratio, $R_r$ of
evolving LMXB and MRP populations, we first emphasize that \ratem~and 
\ratel~are really the instantaneous rates of {\it decay\/} 
of the MRP and LMXB populations, respectively (see eqs. [2]-[3]),
and not their ``birthrates'', as they have been often called in
previous discussions. Only under the assumption of a steady state can 
we equate them to the respective birthrates, and to each other.
For evolving populations, $R_r\approx 1$ is expected only under the 
circumstances described in the last paragraph. Thus, there is no basis 
for expecting $\ratem = \ratel$ in general, and a deviation from 
equality does not, by itself, imply a serious problem. Indeed, since 
the observable quantities are really \mrp~and \lmxb, the actual test 
of agreement is as follows: given a plausible choice of \taup, 
\tauel~and \taum, does the calculated number ratio $N_r$ at $z=0$ at
the present epoch agree with observation, and, furthermore, does the 
calculated rate ratio $R_r$ at the same epoch agree with that 
obtained from the observed $N_r$ with this particular choice of
timescales? 
       
With the discovery of many more MRPs since the original KN work, the 
observational situation has changed somewhat. KN estimated $N_r$ to be 
$\sim 10^2$, and $R_r$ to be $\sim 10$ for the whole population and 
$\sim 100$ for short-period systems. The most recent estimate by L95 
(using $\sim 5$ times as many MRPs as KN) suggests $N_r\approx 400$, 
$R_r\approx 1$ for the whole population, and $R_r\approx 8$ for 
short-period systems. From the case in Figure 1(a), it is clear that 
values of $N_r$ and $R_r$ typically discussed for the whole population 
are naturally obtained in the above picture with canonical timescales 
for the whole population. If short-period systems with longer 
\tauel~(KN; CP) are considered alone, Figure 1(b) shows that $R_r>1$ 
also occurs naturally, but with typical $R_r$ values $\approx 3$. 
It is not clear how significant the discrepancy 
for short-period LMXBs is until a more 
detailed population synthesis has been undertaken. If further work confirms
the discrepancy, we may conclude that either 
(a) a one-to-one evolution from PSNB to LMXB to MRP does not always 
occur: in certain parts of the parameter space, PSNBs do not evolve 
into the LMXB phase, but ultimately produce MRPs (KW98), (b) some 
potential LMXBs are rapidly destroyed, possibly by evaporation of 
the secondary(CP; T91), or, (c) there is a serious undersampling of the LMXB
population because the majority are not X-ray active (L95).        

\section{An Observational Test}

We have demonstrated the inadequacy of steady-state arguments in
discussions of the LMXB-MRP birthrate problem for evolving populations
with a time-variable SFR that peaks at a redshift $\approx 1.5$. We
find that an evolutionary scheme can easily account for the 
observed MRP/LMXB number ratios, \ie, $R_r\approx 1$ for the 
overall population, and larger $R_r$ values for short-period LMXB 
systems. A closely related point is the relative behavior
of high-mass X-ray binary (HMXB) and LMXB populations in a universe
with a time-dependent cosmic SFR. HMXBs and LMXBs originating
from stars formed in the same epoch have very different evolutionary
times, since, although the initial evolution of both involves the
evolution of a massive star to supernova and the formation of a neutron
star, LMXBs turn on as X-ray sources much later than HMXBs, only after
the low-mass companion comes into Roche lobe contact, predominantly
due to orbital decay by gravitational radiation and magnetic braking
(KW98 and references therein). The relevant time lag is essentially the
timescale \taup~introduced in \S 2. Since the post-supernova evolution
into HMXBs takes a negligibly short time on this scale, the global
HMXB population will peak roughly where the number of stars (= 
integral of the SFR) does. Thus, the global LMXB population will peak 
in redshift well after the HMXB peak. The combined 
X-ray binary activity of the two populations is expected to have a 
broad peak, or possibly a double peak, in $z$, depending on the lag 
and the relative population sizes.

The dominant source of X-ray emission from normal spiral galaxies 
(\ie, those without an active nucleus) appears to be the integrated 
emission from their X-ray binaries (see Fabbiano 1995 and references 
herein), based on observations of nearby galaxies such as M31 (where 
individual sources can be resolved) and comparison with the
distribution in our galaxy. These integrated X-ray luminosities are 
in the range $L_x\sim 10^{39}$--$10^{41}$ erg s$^{-1}$ 
and scale linearly with the blue band luminosities of the galaxies.
In our galaxy, LMXB dominate the total X-ray output, and this is 
also the same for M31, where the brightest sources are clustered 
around the bulge. For other relatively nearby galaxies, the average 
X-ray temperatures are in the range $3-6$ keV, also consistent with a
population of LMXB (Kim, Fabbiano \& Trinchieri 1992). HMXB systems 
are also a significant component in some galaxies. They dominate the
1--10 keV output of the irregular LMC and SMC galaxies. The X-ray
outputs of starburst galaxies in the 1-10 keV band seem to be 
dominated by those of their HMXB populations and/or young supernova
remnants (SNR; della Cecca, Griffiths \& Heckman 1997).
 
Our work demonstrates that the peak in the SFR at $z\approx 1.5$ will
cause the integrated X-ray luminosity of galaxies in the redshift range
$0.5-1.0$ to be at least an order of magnitude higher than it is today.
If the current understanding of LMXB evolution is correct, then
a twin-peak signature of the dual LMXB and HMXB-SNR population is 
expected. This is caused by the delayed turn-on of the LMXB population
relative to the short-lived and instantaneous turn-on of the HMXBs
and SNRs associated with the peak in the SFR (Figure 1). 
This second LMXB peak
is in the redshift range $0.5-1.0$ and is caused by the delay of the
secondary in the PSNB to come into contact with its Roche lobe. The
details of this signature, \eg, peak separation, can then be used to
confirm the general picture as to the origin of LMXBs, and to
constrain models for their evolution. The expected flux levels 
($\sim 10^{-15}$--$10^{-16}$ erg cm$^{-2}$ s$^{-1}$) for this 
redshift range will be 
within the capabilities of AXAF, provided sufficient observing time
($\sim 10^5$--$10^6$ s) is dedicated to a suitable field. 
These future observations will provide 
an important new window to understand the evolution of X-ray binaries 
and the resulting millisecond radio pulsar population.

\acknowledgements

It is a pleasure to thank V. Kalogera for communicating results of 
evolutionary calculations in advance of publication, and P. Madau for 
supplying the MVP approximation.  

{}



\clearpage


\begin{figure}[t]
\centerline{\hbox{
\psfig{file=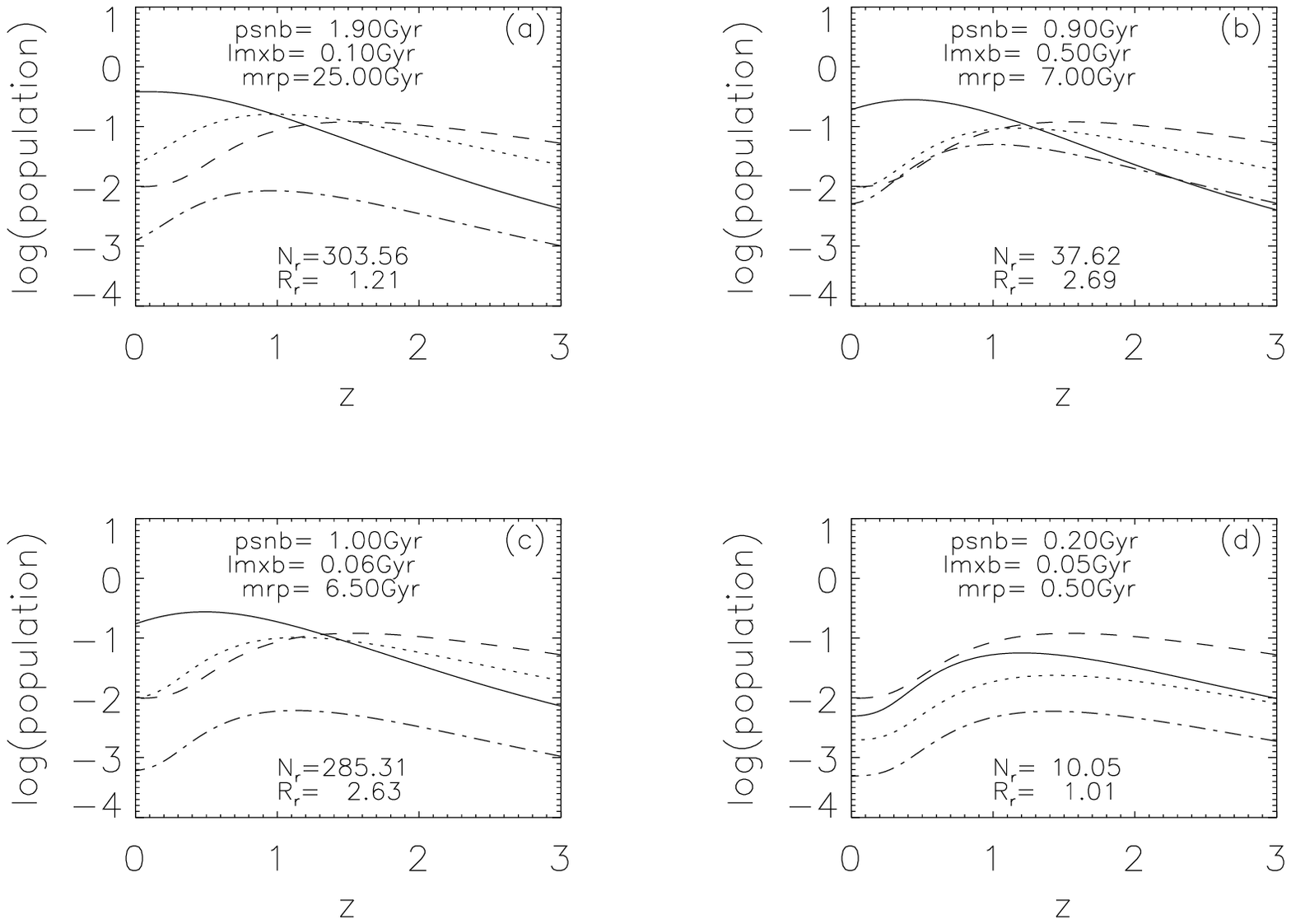,height=10cm,width=12cm}
}}

\caption {Evolution of PSNB, LMXB, and MRP populations in response to a
time-variable cosmic star-formation rate (SFR). Shown are logarithms of 
the number densities of PSNBs (dotted line), LMXBs (dash-dotted line), 
and MRPs (solid line) against the redshift. The SFR of Madau \etal~is
also shown (dashed line) for reference. Each panel displays the
results of an evolutionary calculation with input timescales \taup, 
\tauel, and \taum~written at the top of the panel, and output values
of the number ratio, $N_r\equiv{\mrp\over\lmxb}$, and rate ratio, 
$R_r\equiv{N_{MRP}\over\taum}/{N_{LMXB}\over\tauel}$, at $z=0$ written
at the bottom. 
Case (a) represents a typical result for the whole population of LMXBs
and MRPs, with $R_r\approx 1$. Case (b) represents a typical result 
for short-period systems (see text), with $R_r > 1$. Case (c) shows
the results of postulating an unusually short LMXB lifetime, due, \eg,
to X-ray irradiation of the secondary in close binaries (see text).
Finally, case (d) illustrates the approach to an asymptotic state for
all populations, as described in the text, obtained by choosing 
unrealistically short values for the timescales \taup, \tauel, and 
\taum.} 

\end{figure}

\end{document}